\documentclass[aps,pra,10pt,reprint,floatfix]{revtex4-1}

\usepackage[utf8]{inputenc}
\usepackage[T1]{fontenc}
\usepackage[english]{babel}

\usepackage{graphicx}

\usepackage[version=3]{mhchem}

\usepackage{amsmath}
\usepackage{revsymb}
\usepackage{simpleMath}

\usepackage{units}
\usepackage[squaren]{SIunits}

\usepackage{acronym}

\usepackage{hyperref}

\newcommand{\numw}{\ensuremath{{N_\text{w}}}}
\newcommand{\numg}{\ensuremath{{N_\text{G}}}}
\newcommand{\eff}{\ensuremath{{\text{eff}}}}
\newcommand{\lin}{\ensuremath{{\text{lin}}}}
\newcommand{\trans}{\ensuremath{^\text{T}}}
\newcommand{\nn}{\nonumber}
\newcommand{\pt}{\ensuremath{\mathcal{PT}}}
\newcommand{\erf}{\operatorname{erf}}

\begin{document}

\newacro{BEC}{Bose-Einstein condensate}
\newacro{EOM}{equations of motion}
\newacro{GPE}{Gross-Pitaevskii equation}
\newacro{DNLSE}{discrete nonlinear Schr\"odinger equation}

\title{Realizing \texorpdfstring{$\pt$}{PT}-symmetric non-Hermiticity
  with ultra-cold atoms and Hermitian multi-well potentials}
\author{Manuel Kreibich}
\author{J\"org Main}
\author{Holger Cartarius}
\author{G\"unter Wunner}
\affiliation{Institut f\"ur Theoretische Physik 1, Universit\"at
  Stuttgart, 70550 Stuttgart, Germany}

\begin{abstract}
  We discuss the possibility of realizing a \emph{non-Hermitian},
  i.\,e.\ an \emph{open} two-well system of ultra-cold atoms by
  enclosing it with additional time-dependent wells that serve as
  particle reservoirs. With the appropriate design of the additional
  wells $\mathcal{PT}$-symmetric currents can be induced to and from
  the inner wells, which support stable solutions. We show that
  interaction in the mean-field limit does not destroy this
  property. As a first method we use a simplified variational ansatz
  leading to a discrete nonlinear Schr\"odinger equation. A more
  accurate and more general variational ansatz is then used to confirm
  the results.
\end{abstract}

\maketitle

\section{Introduction}
\label{sec:introduction}

Since the first realization of $\pt$-symmetric gain and loss in
optical wave guides \cite{Rueter10} much effort has been made to
realize analogous systems in various fields of physics, e.\,g.\ lasers
\cite{Chong11}, electronics \cite{Schindler11}, microwave cavities
\cite{Bittner12}, and to make use of new effects arising from
nonlinearity, viz.\ the Kerr nonlinearity in optical wave guides
\cite{Ramezani10}. $\pt$ symmetry originates from quantum mechanics
and stands for a combined action of parity and time-reversal. Despite
the non-Hermiticity of $\pt$-symmetric Hamiltonians, in a certain
range of parameters entirely real eigenvalue spectra exist
\cite{Bender98}. Due to a formal analogy between quantum mechanics and
electromagnetism, the formulation of $\pt$ symmetry has spread to
those systems, leading to the above mentioned realizations. However,
the experimental verification in a genuine quantum system has not been
achieved so far.

In a $\pt$-symmetric system there is a balanced gain and loss of
probability density (or electromagnetic field in analogue
systems). According to a proposal in \cite{Klaiman08} such a quantum
mechanical system could be realized with a \ac{BEC} in a double-well
potential where particles are injected into one well and removed from
the other. Indeed, it could be shown that the system is ideally suited
for a first-experimental observation of $\pt$ symmetry in a quantum
system \cite{Dast13}. There is progress in coupling two \acp{BEC} and
at the same time eject particles \cite{Shin05}, but an actual
realization of $\pt$ symmetry using a \ac{BEC} is still missing.

On the other hand much theoretical work has been done on
$\pt$-symmetric \acp{BEC}, including the proof of existence of stable
states of interacting systems \cite{Cartarius12}, a microscopic
treatment based on the Bose-Hubbard model \cite{Graefe08}, and a
thorough investigation on dynamics of stable and unstable regimes
\cite{Haag14}. In contrast to injecting and removing particles from a
double-well potential, one could think of a double-well included in a
tilted optical lattice, with an incoming and outgoing transport of
particles, so called Wannier-Stark systems
\cite{Zenesini08,Elsen11}. These incoming and outgoing particle
currents could in principle serve as the necessary currents for
realizing a $\pt$-symmetric system.

\begin{figure}
  \centering
  \includegraphics[width=0.5\columnwidth]{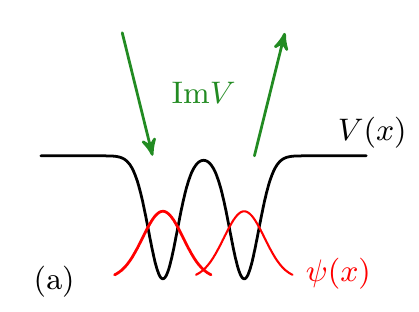}\hfill
  \includegraphics[width=0.5\columnwidth]{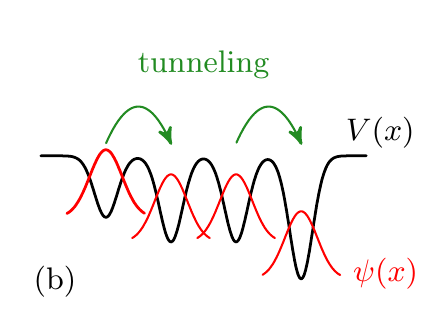}
  \caption{(a) Two-mode model with an imaginary potential $V$, where
    -- in this case -- particles are removed from the left well and
    injected into the right. (b) Such a system could be realized by
    coupling two additional wells to the system, where the tunneling
    currents to and from the outer wells are used as an implementation
    of the imaginary potential. The outer wells act as (limited)
    particle reservoirs and have to be chosen time-dependent.}
  \label{fig:wellcomp}
\end{figure}

In a recent Rapid Communication \cite{Kreibich13} we followed the
simplest approach to this idea, namely to couple two additional wells
to the double well potential (see Fig.~\ref{fig:wellcomp}), and use
the arising currents as a theoretical realization of $\pt$ symmetry in
a quantum mechanical system. We showed that for a specific
time-dependent choice of the well depths and coupling strengths of the
additional wells, the inner wells behave exactly as the wells of a
$\pt$-symmetric two-well system, thus serving as a possible
experimental realization.

It is the purpose of this paper investigate more deeply our approach,
especially for the case of interacting atoms, and to formulate a more
accurate method to confirm the results. In
Sec.~\ref{sec:from-gross-pita} we derive the few-mode model or
\ac{DNLSE} from the \ac{GPE} using a simple variational ansatz, where
each well is populated by a single time-independent Gaussian
function. The results are applied in Sec.~\ref{sec:few-mode-models} to
show that a Hermitian four-mode model can be chosen such that the
middle wells behave exactly as the wells of the $\pt$-symmetric
two-mode model. This is exemplified for some parameters. In
Sec.~\ref{sec:vari-ansatz-with} we use an extended variational ansatz
for the realization of $\pt$ symmetry. Still one Gaussian function per
well is used, but each Gaussian function is fully dynamical in its
parameters, thus giving a more accurate description of the system. The
method is compared with the simply Gaussian ansatz. In
Sec.~\ref{sec:conclusion} we conclude our work.

\section{From the Gross-Pitaevskii equation to a few-mode model}
\label{sec:from-gross-pita}

In this section we apply the steps that lead from the \ac{GPE}, which
describes the dynamics of a \ac{BEC} at absolute zero temperature, to
a \ac{DNLSE} or few-mode-model. At first we presents our ansatz and
calculate the necessary integrals. To derive the \ac{EOM} in a
Schr\"odinger form we use the method of symmetric
orthogonalization. In Sec.~\ref{sec:few-mode-models} the results are
used and applied to different scenarios.

\subsection{Simplified variational ansatz}
\label{sec:simpl-vari-ansatz}

We start with the well-known \ac{GPE},
\begin{align}
  \label{eq:gpe}
  \ii \hbar \partial_t \psi(\vec{r},t) = \left[ -\frac{\hbar^2}{2m}
    \Delta + V(\vec{r}) + g \abs{\psi(\vec{r},t)}^2 \right]
  \psi(\vec{r},t).
\end{align}
The interaction strength is given by $g = 4 \pi \hbar^2 N a / m$ with
the s-wave scattering length $a$ and particle number $N$. For the
external multi-well potential $V(\vec{r})$ we use a Gaussian profile
for each well,
\begin{align}
  \label{eq:pot}
  V(\vec{r}) = \sum\limits_{k=1}^{\numw} V^k \exp \left[ -\frac{2
      x^2}{w_x^2} - \frac{2 y^2}{w_y^2} - \frac{2 (z -
      s_z^k)^2}{w_z^2} \right],
\end{align}
with a total number of $\numw$ wells. Each well $k$ has the potential
depth $V^k < 0$ and is displaced along the $z$-direction by
$s_z^k$. Such a potential could, e.\,g., be created experimentally by
the method presented in \cite{Henderson09}.

There is various work on how to transform the \ac{GPE}~\eqref{eq:gpe}
into a \ac{DNLSE} \cite{Trombettoni01,Smerzi03} by integrating out the
spatial degrees of freedom. In general, an ansatz of the form
\begin{align}
  \label{eq:40}
  \psi(\vec{r},t) = \sum\limits_k d^k(t) g^k(\vec{r}),
\end{align}
is used, where $d^k(t)$ is the (complex) amplitude and $g^k(\vec{r})$
the localized wave function in well $k$. Usually, a set of Wannier
functions is applied \cite{Bloch08}. In this work
[Sec.~\ref{sec:few-mode-models}], we use a single Gaussian function
for each well. With this ansatz we can derive analytical formulae for
the matrix elements of the \ac{DNLSE} or few-mode model, and thus open
an easy way to actually compute the localized wave functions by a
simple energy minimization process. In Sec.~\ref{sec:vari-ansatz-with}
we extend the variational ansatz to obtain more accurate results.

The simplified variational ansatz is given by the superposition of
$\numg$ Gaussian functions, one for each well, $\numw=\numg$, with
\begin{align}
  \label{eq:simpleansatz}
  g^k(\vec{r}) = \eto{-A_x^k x^2 - A_y^k y^2 - A_z^k (z-q_z^k)^2}.
\end{align}
The parameters $A_x^k, A_y^k, A_z^k \in \Reals$ describe the width of
each Gaussian function in each direction, and $q_z^k \in \Reals$ the
displacement along the $z$-direction. We treat these parameters in
this section as time-independent. The time-dependence of the wave
function comes from the amplitude $d^k(t)$ in Eq.~\eqref{eq:40}.

We insert ansatz (\ref{eq:40}) into the \ac{GPE} $\ii \hbar \partial_t
\psi = \hat{H} \psi$, multiply from the left with $(g^l)^*$, and
integrate over $\Reals^3$. Then, we are left with
\begin{align}
  \label{eq:simpleeom}
  \ii \hbar \mat{K} \dot{\vec{d}} = \mat{H} \vec{d}.
\end{align}
Here, the amplitudes $d^k$ are written as a vector $\vec{d} = (d^1,
\dots, d^\numg)^\text{T} \in \Complex^\numg$. The matrix $\mat{K}$
describes the overlap, and $\mat{H}$ the Hamiltonian matrix
elements. They are given by the integrals (see
Sec.~\ref{sec:eval-expr} for their explicit calculation)
\begin{subequations}
  \label{eq:46}
  \begin{align}
    K_{lk} &= \qmprod{g^l}{g^k}, \\
    H_{lk} &= \matrixel{g^l}{\hat{H}}{g^k}.
  \end{align}
\end{subequations}
Our aim is to derive \ac{EOM} for the amplitudes in a Schr\"odinger
form, i.\,e. $\ii \hbar \dot{\vec{d}}_\eff = \mat{H}_\eff
\vec{d}_\eff$, with possibly modified amplitudes and matrix
elements. Equation~\eqref{eq:simpleeom} would be of that form if
$\mat{K}$ was proportional to the identity matrix. Since the Gaussian
functions $g^k$ are non-orthogonal this is not the case. For that
reason, we perform a symmetric orthogonalization in
Sec.~\ref{sec:symm-orth}.

\subsection{Evaluation of the integrals}
\label{sec:eval-expr}

The occurring integrals in Eqs.~\eqref{eq:46} are solely based on the
Gaussian integral
\begin{align}
  \label{eq:gaussint}
  \int\limits_{-\infty}^\infty \! \dd x \, x^n \eto{-A x^2 + p x} &=
  \frac{\partial^n}{\partial p^n} \int\limits_{-\infty}^\infty
  \! \dd x \, \eto{-A x^2 + p x} \nn\\
  &= \frac{\partial^n}{\partial p^n} \sqrt{\frac{\pi}{A}} \eto{p^2/4
    A}, \qquad \Real A > 0.
\end{align}
For convenience we define the abbreviations
\begin{subequations}
  \begin{align}
    \label{eq:5}
    A_\alpha^{kl} &= A_\alpha^k + (A_\alpha^l)^*, \qquad \alpha =
    x,y,z, \\
    \kappa_\alpha^{kl} &= A_\alpha^k (A_\alpha^l)^* / A_\alpha^{kl}, \\
    \beta_\alpha^{kl} &= \sqrt{\frac{A_\alpha^{kl}
        w_\alpha^2}{A_\alpha^{kl} w_\alpha^2 + 2}}, \\
    c^{kl} &= \eto{-\kappa_z^{kl} (q_z^k - q_z^l)^2}.
  \end{align}  
\end{subequations}
The integrals of $\mat{K}$ can now easily be calculated with
Eq.~\eqref{eq:gaussint} for the case $n=0$, the result is
\begin{align}
  \label{eq:6}
  K_{lk} = \sqrt{\frac{\pi}{A_x^{kl}}} \sqrt{\frac{\pi}{A_y^{kl}}}
  \sqrt{\frac{\pi}{A_z^{kl}}} c^{kl}.
\end{align}
For the calculation of the Hamiltonian matrix $\mat{H}$, we separate
it into a kinetic term $\mat{T}$, an external potential term
$\mat{V}$, and an interaction term $\mat{W}$.

For the kinetic term we need integrals of the form shown in
Eq.~\eqref{eq:gaussint} with $n=2$ and we obtain
\begin{align}
  \label{eq:intkin}
  T_{lk} = K_{lk} \left[ \kappa_x^{kl} + \kappa_y^{kl} + \kappa_z^{kl}
    - 2 (\kappa_z^{kl})^2 (q_z^k - q_z^l)^2 \right].
\end{align}
To calculate the integrals necessary for the external potential, the
parameters $A$ and $p$ in Eq.~\eqref{eq:gaussint} are shifted by the
parameters of the external potential. The result is
\begin{multline}
  \label{eq:intpot}
  V_{lk} = K_{lk} \beta_x^{kl} \beta_y^{kl} \beta_z^{kl} \sum\limits_m
  V^m \\
  \times \exp \left\{ -\frac{2 \left[ A_z^k (s_z^m - q_z^k) +
        (A_z^l)^* (s_z^m - q_z^l) \right]^2}{A_z^{kl} (A_z^{kl} w_z^2
      + 2)} \right\}.
\end{multline}
Since the operator of the interaction potential $\mat{W}$ is
nonlinear, its matrix elements depend on the amplitudes $d^k$. With
the standard Gaussian integral we obtain
\begin{align}
  \label{eq:4}
  W_{lk} = \sum\limits_{i,j} \tilde{W}_{lkji} (d^j)^* d^i,
\end{align}
where we defined the four-rank tensor
\begin{multline}
  \label{eq:9}
  \tilde{W}_{lkji} = \frac{4 \pi \hbar^2 N a}{m}
  \sqrt{\frac{\pi}{A_x^{ijkl}}} \sqrt{\frac{\pi}{A_y^{ijkl}}}
  \sqrt{\frac{\pi}{A_z^{ijkl}}} \\
  \times \exp \left[ -\frac{A_z^i (A_z^j)^* (q_z^i - q_z^j)^2 + A_z^i
      (A_z^l)^* (q_z^i - q_z^l)^2}{A_z^{ijkl}} \right] \\
  \times \exp \left[ -\frac{A_z^k (A_z^j)^* (q_z^k - q_z^j)^2 +
      A_z^k (A_z^l)^* (q_z^k - q_z^l)^2}{A_z^{ijkl}} \right] \\
  \times \exp \left[ -\frac{A_z^i A_z^k (q_z^i - q_z^k)^2 + (A_z^j)^*
      (A_z^l)^* (q_z^j - q_z^l)^2 }{A_z^{ijkl}} \right]
\end{multline}
and the new abbreviations
\begin{align}
  \label{eq:10}
  A_\alpha^{ijkl} = A_\alpha^{ij} + A_\alpha^{kl}, \qquad \alpha =
  x,y,z.
\end{align}

\subsection{Symmetric orthogonalization}
\label{sec:symm-orth}

We calculated all necessary integrals for setting up the \acp{EOM}
\eqref{eq:simpleeom}, which are not of the form of a Schr\"odinger
equation (cf.\ Sec.~\ref{sec:simpl-vari-ansatz}). To achieve this, we
now transform the \ac{EOM} into a Schr\"odinger form with the method
of symmetric orthogonalization \cite{Lowdin50}: Since $\mat{K}$ is
Hermitian there exists a unitary transformation $\mat{U}$ such that
$\mat{D} = \mat{U} \mat{K} \mat{U}^\dagger$ is diagonal with real
entries. Then we can construct the Hermitian matrix
\begin{align}
  \label{eq:xdef}
  \mat{X} = \mat{U}^\dagger \mat{D}^{-1/2} \mat{U}.
\end{align}
This matrix has the properties that $\mat{X} \mat{K} \mat{X} =
\identity$, and furthermore, if $\mat{H}$ is Hermitian, so is $\mat{X}
\mat{H} \mat{X}$. Now, we can use this matrix to transform
Eq.~\eqref{eq:simpleeom} to
\begin{align}
  \label{eq:1}
  \ii \hbar (\mat{X} \mat{K} \mat{X}) (\mat{X}^{-1} \dot{\vec{d}}) =
  (\mat{X} \mat{H} \mat{X}) (\mat{X}^{-1} \vec{d}).
\end{align}
With the definitions $\vec{d}_\eff = \mat{X}^{-1} \vec{d}$ and
$\mat{H}_\eff = \mat{X} \mat{H} \mat{X}$, and the above property of
$\mat{X}$ we arrive at the \ac{EOM} for $\vec{d}_\eff$ in
Schr\"odinger form
\begin{align}
  \label{eq:2}
  \ii \hbar \dot{\vec{d}}_\eff = \mat{H}_\eff \vec{d}_\eff
\end{align}
with a Hermitian Hamiltonian matrix $\mat{H}_\eff$ (if $\mat{H}$ is
Hermitian). We note, that the normalization condition for the wave
function $\psi$ transforms as
\begin{align}
  \label{eq:norm}
  1 &\stackrel{!}{=} \intdthreer \abs{\psi}^2 = \vec{d}^\dagger \mat{K}
  \vec{d} = (\mat{X}^{-1} \vec{d})^\dagger (\mat{X} \mat{K} \mat{X})
  (\mat{X}^{-1} \vec{d}) \nn\\
  &= \vec{d}_\eff^\dagger \vec{d}_\eff.
\end{align}

\subsection{Application of symmetric orthogonalization and
  nearest-neighbor approximation}
\label{sec:appl-symm-orth}

By means of symmetric orthogonalization we can transform the \ac{EOM}
\eqref{eq:simpleeom} into a Schr\"odinger form since we know $\mat{K}$
analytically. Without any further approximations the analytical
expressions can become quite complicated. At this point, we introduce
as usual the nearest neighbor approximation to obtain analytical and
simple approximate expressions for $\mat{X}$ and $\mat{H}_\eff$.

The structure of the matrix elements $K_{lk}$ [see Eq.~\eqref{eq:6}]
allows for a natural approximation since the function $c^{kl}$ drops
exponentially with increasing distance of wells $\abs{k-l}$. We say a
term is of order $n$ when $\abs{k-l}=n$. As an approximation we only
consider terms up to order $n=1$. Hence, we have $\mat{K} =
\mat{K}^{(0)} + \mat{K}^{(1)}$ with
\begin{align}
  \label{eq:11}
  K_{lk}^{(0)} = \left( \frac{\pi}{2} \right)^{3/2}
  \frac{\delta_{kl}}{\sqrt{A_{x,R}^k A_{y,R}^k A_{z,R}^k}}
\end{align}
and
\begin{align}
  \label{eq:12}
  K_{lk}^{(1)} = \sqrt{\frac{\pi}{A_x^{kl}}}
  \sqrt{\frac{\pi}{A_y^{kl}}} \sqrt{\frac{\pi}{A_z^{kl}}} c^{kl}
  \left( \delta_{k,l+1} + \delta_{k+1,l} \right),
\end{align}
where $A_{\alpha,R}^k = \Real A_\alpha^k$, and -- for later use --
$A_{\alpha,I}^k = \Imag A_\alpha^k$.

Using this expansion, we calculate the zeroth order of $\mat{X}$ by
the requirement $\mat{X}^{(0)} \mat{K}^{(0)} \mat{X}^{(0)} =
\identity$. We obtain
\begin{align}
  \label{eq:13}
  X_{lk}^{(0)} = \left( \frac{2}{\pi} \right)^{3/4} \sqrt[4]{A_{x,R}^k
    A_{y,R}^k A_{z,R}^k} \delta_{kl}.
\end{align}
To calculate the first order we start with
\begin{align}
  \label{eq:14}
  \left( \mat{X}^{(0)} + \mat{X}^{(1)} \right) \left( \mat{K}^{(0)} +
    \mat{K}^{(1)} \right) \left( \mat{X}^{(0)} + \mat{X}^{(1)} \right)
  = \identity,
\end{align}
ignore second order terms, insert the known quantities, and solve for
$\mat{X}^{(1)}$, which yields
\begin{multline}
  \label{eq:15}
  X_{lk}^{(1)} = - \left( \frac{8}{\pi} \right)^{3/4}
  \frac{c^{kl}}{\sqrt{A_x^{kl}} \sqrt{A_y^{kl}} \sqrt{A_z^{kl}}}
  \left( \delta_{k,l+1} + \delta_{k+1,l} \right) \\
  \times \frac{\sqrt{A_{x,R}^k A_{y,R}^k A_{z,R}^k A_{x,R}^l A_{y,R}^l
      A_{z,R}^l}}{\sqrt[4]{A_{x,R}^k A_{y,R}^k A_{z,R}^k} +
    \sqrt[4]{A_{x,R}^l A_{y,R}^l A_{z,R}^l}}.
\end{multline}

Now we are able to calculate analytical approximations for the
transformed Hamiltonian $\mat{H}_\eff$. We observe that for the linear
parts, i.\,e.\ the kinetic and external potential part, the matrix
elements of $\mat{H}$ are proportional to $\mat{K}$, so we can write
$H_{lk}^\lin = K_{lk} h_{lk}^\lin$ [cf. Eqs.~\eqref{eq:intkin}
and~\eqref{eq:intpot}]. Thus, the orders of $\mat{H}^\lin$ are given
by the orders of $\mat{K}$. Therefore, for the zeroth order
transformed Hamiltonian we obtain
\begin{align}
  \label{eq:16}
  H_{\eff,lk}^{\lin,(0)} = \sum\limits_{m,n} X_{lm}^{(0)} K_{mn}^{(0)}
  X_{nk}^{(0)} h_{mn}^\lin = h_{kk}^\lin \delta_{kl}.
\end{align}
The zeroth order contributes to the linear diagonal elements of
$\mat{H}_\eff^\lin$, so they can be identified with the onsite energy
$E_k = h_{kk}^\lin$. With Eqs.~\eqref{eq:intkin} and~\eqref{eq:intpot}
we can write
\begin{multline}
  \label{eq:7}
  E_k = \frac{\hbar^2}{2 m} \left( A_{x,R}^k + A_{y,R}^k + A_{z,R}^k
  \right) \\
  + V^k \beta_x^{kk} \beta_y^{kk} \beta_z^{kk} \exp \left[ -2
    (\beta_z^{kk})^2 (s_z^k - q_z^k)^2 / w_z^2 \right],
\end{multline}
where we considered only the term with $m=k$ in the sum in
Eq.~\eqref{eq:intpot} to be consistent with zeroth order.

The first order terms of $\mat{H}_\eff^\lin$ are given by the first
order terms of the expression
\begin{align}
  \label{eq:8}
  (\mat{X}^{(0)} + \mat{X}^{(1)}) (\mat{H}^{\lin,(0)} +
  \mat{H}^{\lin,(1)}) (\mat{X}^{(0)} + \mat{X}^{(1)}),
\end{align}
which gives, after inserting all known quantities,
\begin{align}
  \label{eq:17}
  H_{\eff,lk}^{\lin,(1)} = &- 2 \sqrt{2} \frac{\sqrt{A_{x,R}^k
      A_{y,R}^k A_{z,R}^k A_{x,R}^l A_{y,R}^l
      A_{z,R}^l}}{\sqrt[4]{A_{x,R}^k A_{y,R}^k A_{z,R}^k} +
    \sqrt[4]{A_{x,R}^l A_{y,R}^l A_{z,R}^l}} \nn\\
  &\times \left( \frac{E_k - h_{lk}}{\sqrt[4]{A_{x,R}^k A_{y,R}^k
        A_{z,R}^k}} + \frac{E_l - h_{lk}}{\sqrt[4]{A_{x,R}^l A_{y,R}^l
        A_{z,R}^l}} \right) \nn\\
  &\times \frac{c^{kl}}{\sqrt{A_x^{kl}} \sqrt{A_y^{kl}}
    \sqrt{A_z^{kl}}} \left( \delta_{k,l+1} + \delta_{k+1,l} \right).
\end{align}
These elements can be identified as the tunneling elements of the
few-mode model, i.\,e.\ $J_{lk} = -H_{\eff,lk}^{\lin,(1)}$, since they
are the super- and sub-diagonal entries. The quantity $h_{lk}^\lin$ is
given by Eqs.~\eqref{eq:intkin} and~\eqref{eq:intpot}, where in
$v_{lk}$ we only consider nearest-neighbor contributions,
\begin{multline}
  \label{eq:18}
  v_{lk} = \beta_x^{kl} \beta_y^{kl} \beta_z^{kl} \\
  \times \Biggl( V^k \exp \left\{ -\frac{2 \left[ A_z^k (s_z^k -
        q_z^k) + (A_z^l)^* (s_z^k - q_z^l) \right]^2}{A_z^{kl}
      (A_z^{kl} w_z^2 + 2)} \right\} \\
  + V^l \exp \left\{ -\frac{2 \left[ A_z^k (s_z^l - q_z^k) + (A_z^l)^*
        (s_z^l - q_z^l) \right]^2}{A_z^{kl} (A_z^{kl} w_z^2 + 2)}
  \right\} \Biggr).
\end{multline}

So far we have calculated the matrix elements of the linear
transformed Hamiltonian $\mat{H}_\eff^\lin$. We now turn our attention
to the interaction part. Since the operator $\mat{W}$ is nonlinear and
depends on the amplitudes $d^k$, they must be transformed, too. For
the transformed operator, we obtain
\begin{align}
  \label{eq:19}
  W_{\eff,lk} &= \sum\limits_{i,j,m,n,p,q} X_{lm} \tilde{W}_{mnji}
  X_{nk} (X_{jp} d_\eff^p)^* (X_{iq} d_\eff^q) \nn\\
  &= \sum\limits_{p,q} \underbrace{\sum\limits_{i,j,m,n} X_{lm} X_{pj}
    \tilde{W}_{mnji} X_{nk} X_{iq}}_{\equiv \tilde{W}_{\eff,lkpq}}
  (d_\eff^p)^* d_\eff^q.
\end{align}
For the interaction term, we only consider onsite-elements, i.\,e.\
elements with $i=j=k=l$ in $\tilde{W}_{lkji}$, since mainly these
terms contribute to the interaction part. For the transformed
four-rank tensor, we then obtain
\begin{align}
  \label{eq:20}
  \tilde{W}_{\eff,kkkk} = \frac{4 \hbar^2 N a}{\sqrt{\pi} m}
  \sqrt{A_{x,R}^k A_{y,R}^k A_{z,R}^k}.
\end{align}
These elements are the nonlinear coupling constants, $c_k =
\tilde{W}_{\eff,kkkk}$. The whole action of the transformed
Hamiltonian then reads
\begin{align}
  \label{eq:21}
  W_{\eff,lk} = c_k |d^k|^2 \delta_{kl}.
\end{align}

We have all necessary matrix-elements in nearest-neighbor
approximation. Using this knowledge we can calculate these elements
from the Gaussian variational approach, or -- vice verse -- determine
the parameters of the realistic potential \eqref{eq:pot} by knowing
the matrix elements. As an application this is done in
Sec.~\ref{sec:few-mode-models}. The parameters $A_\alpha^k$ and
$q_z^k$ can be computed by minimizing the mean-field energy of the
system for a given initial condition with
\begin{multline}
  \label{eq:22}
  E_\text{mf} = \sum\limits_{k,l} (d^l)^* \left( T_{lk} + V_{lk}
  \right) d^k \\
  + \frac{1}{2} \sum\limits_{i,j,k,l} (d^l)^* d^k W_{lkji} (d^j)^*
  d^i.
\end{multline}
For this minimization, all parameters $A_\alpha^k$, $q_z^k$ and $d^k$
have to be varied with the norm constraint \eqref{eq:norm}, which is
computationally much cheaper than a grid calculation. The results of
this section could also be used to calculate the parameters of the
Bose-Hubbard model.

Now that we have all matrix elements, we can begin our actual
investigation of a realization of a non-Hermitian system, first by
means of few-mode models. In the following we will adopt the usual
notation and write for the discrete wave function $\psi_k$ instead of
$d^k$.

\section{Few-Mode models}
\label{sec:few-mode-models}

\subsection{Equivalence of the two- and four-mode models}
\label{sec:equivalence-two-four}

We now return to the idea mentioned in the introduction and sketched
in Fig.~\ref{fig:wellcomp}, i.\,e.\ the embedding of a non-Hermitian
double-well with a closed four-well structure. Before analyzing the
possibility of realizing a non-Hermitian system with a Hermitian
model, we discuss the basic properties of the non-Hermitian two-mode
model, which is given by
\begin{align}
  \label{eq:ham2mode}
  \mat{H}_2 =
  \begin{pmatrix}
    E_1 + c_1 \abs{\psi_1}^2 & -J_{12} \\
    -J_{12} & E_2 + c_2 \abs{\psi_2}^2
  \end{pmatrix}
\end{align}
with complex onsite energies $E_k \in \Complex$, which make the
Hamiltonian non-Hermitian. This system has intensively been
investigated in \cite{Graefe12}. The time derivatives of the
observables, that is particle number $n_k = \psi_k^* \psi_k$ and
particle current $j_{kl} = \ii J_{kl} ( \psi_k \psi_l^* - \psi_k^*
\psi_l)$, can be easily calculated, which gives
\begin{subequations}
  \label{eq:2modedt}
  \begin{align}
    \label{eq:2modedt1}
    \partial_t n_1 &= -j_{12} + 2 n_1 \Imag E_1, \\
    \label{eq:2modedt2}
    \partial_t n_2 &= j_{12} + 2 n_2 \Imag E_2, \\
    \label{eq:2modedt3}
    \partial_t j_{12} &= 2 J_{12}^2 (n_1 - n_2) + (\Imag E_1 + \Imag
    E_2) j_{12} \nn\\
    &\quad + J_{12} ( \Real E_1 - \Real E_2 + c_1 n_1 - c_2 n_2 )
    C_{12}, \\
    \label{eq:2modedt4}
    \partial_t C_{12} &= (\Imag E_1 + \Imag E_2) C_{12} \nn\\
    &\quad - ( \Real E_1 - \Real E_2 + c_1 n_1 - c_2 n_2 )
    \tilde{j}_{12},
  \end{align}
\end{subequations}
where we defined $C_{kl} = \psi_k \psi_l + \psi_k^* \psi_l^*$ and
$\tilde{j}_{kl} = \ii ( \psi_k \psi_l^* - \psi_k^* \psi_l) = j_{12} /
J_{12}$. In Eqs.~\eqref{eq:2modedt1} and~\eqref{eq:2modedt2} we see
that the imaginary parts of the onsite energies act as additional
sources or sinks to the particle flow, the quantities $j_{e1} = 2 n_1
\Imag E_1$ and $j_{e2} = - 2 n_2 \Imag E_2$ can be identified as
currents to and from the environment, thus, this Hamiltonian describes
an open quantum system.

The general solutions for the nonlinear model \eqref{eq:ham2mode} are
calculated in \cite{Graefe12}. For convinience we only discuss the
results of the linear, i.\,e.\ non-interacting case ($c_1 = c_2 =
0$). Then, the eigenvalues are given by
\begin{align}
  \label{eq:3}
  E_\pm = \frac{E_1+E_2}{2} \pm \sqrt{J_{12}^2 +
    \frac{(E_1-E_2)^2}{2}}.
\end{align}
Requirement for $\pt$ symmetry leads to the condition
$E_1 = E_2^*$, which is for instance fulfilled by $E_1 = \ii \Gamma$
and $E_2 = -\ii \Gamma$, a situation with balanced gain and
loss. Then, the eigenvalues are $E_\pm = \sqrt{J_{12}^2 -
  \Gamma^2}$. For $\Gamma < J_{12}$ they are real, whereas for $\Gamma
> J_{12}$ the $\pt$ symmetry is broken and the eigenvalues are purely
imaginary.

It is now our main purpose to investigate whether the behavior of the
non-Hermitian two-mode model \eqref{eq:ham2mode} can be described by a
Hermitian four-mode model, which is given by
\begin{multline}
  \label{eq:23}
  \mat{H}_4(t) =
  \begin{pmatrix}
    E_0(t) & -J_{01}(t) & 0 & 0 \\
    -J_{01}(t) & E_1 & -J_{12} & 0 \\
    0 & -J_{12} & E_2 & -J_{23}(t) \\
    0 & 0 & -J_{23}(t) & E_3(t)
  \end{pmatrix} \\
  + \diag \left( c_0 \abs{\psi_0}^2, c_1 \abs{\psi_1}^2, c_2
    \abs{\psi_2}^2, c_3 \abs{\psi_3}^2 \right).
\end{multline}
Two additional wells are coupled to the system, they act as a particle
reservoir for the inner wells. The onsite energies and tunneling
elements of the outer wells may be time-dependent. Since this model
shall be Hermitian, the onsite energies are real, $E_k \in \Reals$.

To derive a relationship between the two- and four-mode model we
calculate the time derivatives of the same observables as in the
two-mode model, which yields
\begin{subequations}
  \label{eq:4modedt}
  \begin{align}
    \partial_t n_1 &= j_{01} - j_{12}, \\
    \partial_t n_2 &= j_{12} - j_{23}, \\
    \partial_t j_{12} &= 2 J_{12}^2 \left( n_1 - n_2 \right) - J_{12}
    \left( J_{01} C_{02} - J_{23} C_{13} \right) \nn\\
    &\quad + J_{12} \left( E_1 - E_2 + c_1 n_1 - c_2 n_2 \right)
    C_{12}, \\
    \partial_t C_{12} &= J_{01} \tilde{j}_{02} - J_{23} \tilde{j}_{13}
    - \left( E_1 - E_2 + c_1 n_1 - c_2 n_2 \right) \tilde{j}_{12}.
  \end{align}
\end{subequations}
Comparing with Eqs.~\eqref{eq:2modedt} we find that for the two-mode
model the following condition must hold,
\begin{align}
  \label{eq:26}
  \Imag E_1 = -\Imag E_2 \Rightarrow E_1 = E_2^*.
\end{align}
This means that only the $\pt$-symmetric two-mode model, as discussed
above, may be simulated by the four-mode model. From now on we write
$E_1 = \ii \Gamma$ and $E_2 = -\ii \Gamma$ with $\Gamma \in
\Reals$. Furthermore, the real parts of the onsite energies $E_1$ and
$E_2$ and the coupling element $J_{12}$ have to agree between the two
models. We are then left with the following conditions,
\begin{subequations}
  \label{eq:cond}
  \begin{align}
    \label{eq:cond1}
    j_{01} &= 2 \Gamma n_1, \\
    \label{eq:cond2}
    j_{23} &= 2 \Gamma n_2, \\
    \label{eq:cond3}
    0 &= J_{01} C_{02} - J_{23} C_{13}, \\
    \label{eq:cond4}
    0 &= J_{01} \tilde{j}_{02} - J_{23} \tilde{j}_{13}.
  \end{align}
\end{subequations}
To summarize at this point, if these conditions are fulfilled at every
time, the two middle wells of the Hermitian four-mode model behave
exactly as the wells of the $\pt$-symmetric two-mode model. These four
conditions seem to be independent, but we shall see in
App.~\ref{sec:analyt-cons} that Eq.~\eqref{eq:cond4} follows if
Eqs.~\eqref{eq:cond1}--\eqref{eq:cond3} are fulfilled.

We now have to make clear that these conditions can indeed be
fulfilled by giving suitable time-dependencies to the elements of the
four-mode Hamiltonian. Equation~\eqref{eq:cond3} can simply be
fulfilled by choosing
\begin{align}
  \label{eq:jsol}
  J_{01}(t) = d C_{13}(t), && J_{23}(t) = d C_{02}(t),
\end{align}
where $d \neq 0$ is a real, time-independent quantity. The tunneling
elements are then time-dependent. With $J_{01}$ and $J_{23}$
determined there are no free parameters left to fulfill
Eqs.~\eqref{eq:cond1} and~\eqref{eq:cond2}. Instead, we can set the
time-derivatives of these equations with the help of the remaining
onsite energies $E_0$ and $E_3$. We calculate the time-derivatives of
$j_{01}$ and $j_{23}$ using the Schr\"odinger equation, which yields
\begin{subequations}
  \label{eq:currdt}
  \begin{align}
    \partial_tj_{01} &= \left( \partial_t J_{01} \right)
    \tilde{j}_{01} + 2 J_{01}^2 \left( n_0 - n_1 \right) + J_{01}
    J_{12} C_{02} \nn\\
    &\quad + J_{01} \left( E_{0} - E_{1} + c_0 n_0 - c_1 n_1 \right)
    C_{01} \nn\\
    &\stackrel{!}{=} \partial_t j_{01}^\text{tar}, \\
    \partial_tj_{23} &= \left( \partial_t J_{23} \right)
    \tilde{j}_{23} + 2 J_{23}^2 \left( n_2 - n_3 \right) - J_{23}
    J_{12} C_{13} \nn\\
    &\quad + J_{23} \left( E_{2} - E_{3} + c_2 n_2 - c_3 n_3 \right)
    C_{23} \nn\\
    &\stackrel{!}{=} \partial_t j_{23}^\text{tar},
  \end{align}
\end{subequations}
with the target currents determined by Eqs.~\eqref{eq:cond1}
and~\eqref{eq:cond2}. The time derivatives of the tunneling elements
$J_{01}$ and $J_{23}$ are given by
\begin{subequations}
  \begin{align}
    \label{eq:28}
    \partial_t J_{01} &= d \partial_t C_{13} \nn\\
    &= d \left( J_{01} \tilde{j}_{03} + J_{12} \tilde{j}_{23} - J_{23}
      \tilde{j}_{12} \right) \nn\\
    &\quad + d \left( E_3 - E_1 + c_3 n_3 - c_1 n_1 \right)
    \tilde{j}_{13}, \\
    \partial_t J_{23} &= d \partial_t C_{02} \nn\\
    &= d \left( J_{01} \tilde{j}_{12} - J_{12} \tilde{j}_{01} - J_{23}
      \tilde{j}_{03} \right) \nn\\
    &\quad + d \left( E_2 - E_0 + c_2 n_2 - c_0 n_0 \right)
    \tilde{j}_{02}.
  \end{align}
\end{subequations}
Inserting this into Eqs.~\eqref{eq:currdt}, the onsite energies $E_0$
and $E_3$ are determined by a linear set of equations,
\begin{align}
  \label{eq:lgs}
  \begin{pmatrix}
    C_{01} C_{13} & \tilde{j}_{01} \tilde{j}_{13} \\
    -\tilde{j}_{02} \tilde{j}_{23} & -C_{02} C_{23}
  \end{pmatrix}
  \begin{pmatrix}
    E_0 \\ E_3
  \end{pmatrix}
  =
  \begin{pmatrix}
    v_0 \\ v_3
  \end{pmatrix},
\end{align}
with the entries
\begin{subequations}
  \begin{align}
    \label{eq:29}
    v_0 &= \partial_t j_{01}^\text{tar} / d - J_{12} \tilde{j}_{23} -
    d \left( C_{13} \tilde{j}_{03} - C_{02} \tilde{j}_{12}
    \right) \tilde{j}_{01} \nn\\
    &\quad - 2 d C_{13}^2 \left( n_0 - n_1 \right) - J_{12}
    C_{02} C_{13} \nn\\
    &\quad - \left( c_3 n_3 - c_1 n_1 \right) \tilde{j}_{01}
    \tilde{j}_{13} - \left( c_0 n_0 - c_1 n_1 \right) C_{01} C_{13},
    \\
    v_3 &= \partial_t j_{23}^\text{tar} / d + J_{12} \tilde{j}_{01} -
    d \left( C_{13} \tilde{j}_{12} - C_{02} \tilde{j}_{03}
    \right) \tilde{j}_{23} \nn\\
    &\quad - 2 d C_{02}^2 \left( n_2 - n_3 \right) + J_{12}
    C_{02} C_{13} \nn\\
    &\quad - \left( c_2 n_2 - c_0 n_0 \right) \tilde{j}_{02}
    \tilde{j}_{23} - \left( c_2 n_2 - c_3 n_3 \right) C_{02} C_{23},
  \end{align}
\end{subequations}
where we set $E_1 = E_2 = 0$, since -- as discussed above -- the real
parts of these elements have to agree between the two- and four-mode
model.

We have to note that by means of the onsite energies $E_0$ and $E_3$
we do not fix the currents, but their time-derivatives. For this
reason the initial probabilities and currents have to be chosen such
that they fulfill the conditions~\eqref{eq:cond}. In the following
section we apply these results to obtain solutions for different
starting conditions.

\subsection{Results}
\label{sec:results}

Before presenting our results we give the necessary initial
conditions. We insert the definition of the particle current and the
solutions for the tunneling elements~\eqref{eq:jsol} into the two
Eqs.~\eqref{eq:cond1} and~\eqref{eq:cond2} and express everything in
terms of the wave function $\psi_k$. We are allowed to choose the
global phase of the initial wave function, so we choose $\psi_2^I =
0$. We assume the real parts $\psi_0^R$ and $\psi_3^R$ to be arbitrary
but fixed. We are left with (after simple rearrangement)
\begin{subequations}
  \begin{align}
    \label{eq:25}
    \psi_3^I &= \frac{\Gamma}{2 d \psi_0^R}, \\
    \psi_0^I &= \frac{\psi_0^R \psi_1^I}{\psi_1^R} - \frac{\Gamma
      n_1}{2 d \psi_1^R \left( \psi_1^R \psi_3^R + \psi_1^I \psi_3^I
      \right)}.
  \end{align}
\end{subequations}
Now, the wave function $\psi_k$ fulfills the necessary conditions and
with the time-dependencies of the matrix elements from
Sec.~\ref{sec:equivalence-two-four} we can simulate the
$\pt$-symmetric two-mode model.

\begin{figure}[tb]
  \centering
  \includegraphics[width=\columnwidth]{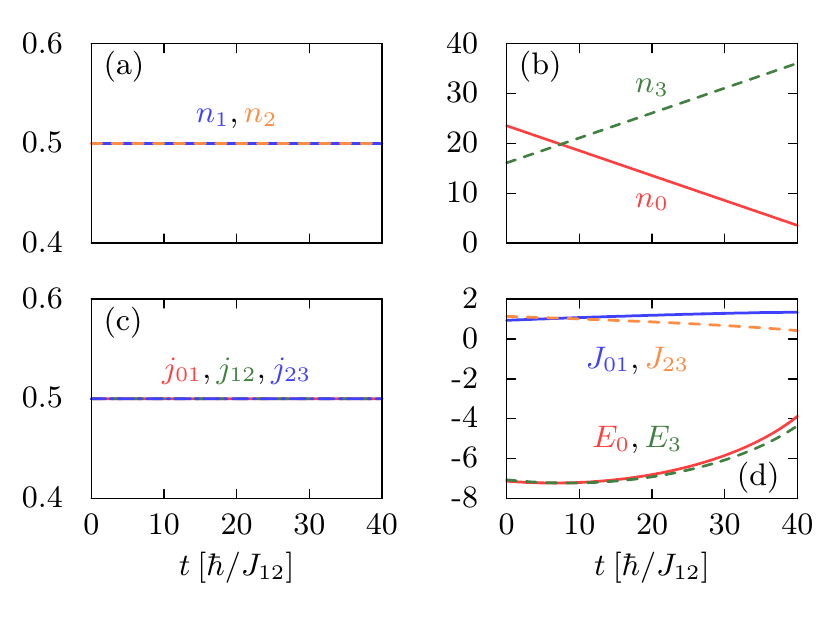}
  \caption{Realization of a stationary $\pt$-symmetric solution within
    the four-mode model. (a) Particles numbers in well~1 (solid) and
    well~2 (dashed), both equal to $1/2$. (b) Particle numbers in
    well~0 (solid) and well~3 (dashed). (c) Particle currents $j_{01}$
    (solid), $j_{12}$ (dashed) and $j_{23}$ (dash-dotted) in units of
    $J_{12}/\hbar$. (d) Tunneling elements $J_{01}$ (solid), $J_{23}$
    (dashed), and onsite energies $E_0$ (solid), $E_3$ (dashed), in
    units of $J_{12}$.}
  \label{fig:stat_sol}
\end{figure}

First we consider a stationary solution in the non-interacting case
[cf.\ Eq.~\eqref{eq:3}] for $\Gamma/J_{12} =
0.5$. Fig.~\ref{fig:stat_sol} shows the results. As expected the
particle numbers in the two middle wells are both equal and constant
in time (with normalization $n_1+n_2=1$), as well as the particle
currents. For this reason, the number of particles decreases (left
well) or increases linearly (right well), with the slope $\dot{n}_0 =
-\Gamma/\hbar$ and $\dot{n}_3 = \Gamma/\hbar$, respectively. The
matrix elements of the four-mode model vary slightly in time. Thus, we
have shown that the conditions~\eqref{eq:cond} can be fulfilled for a
finite time.

\begin{figure}[tb]
  \centering
  \includegraphics[width=\columnwidth]{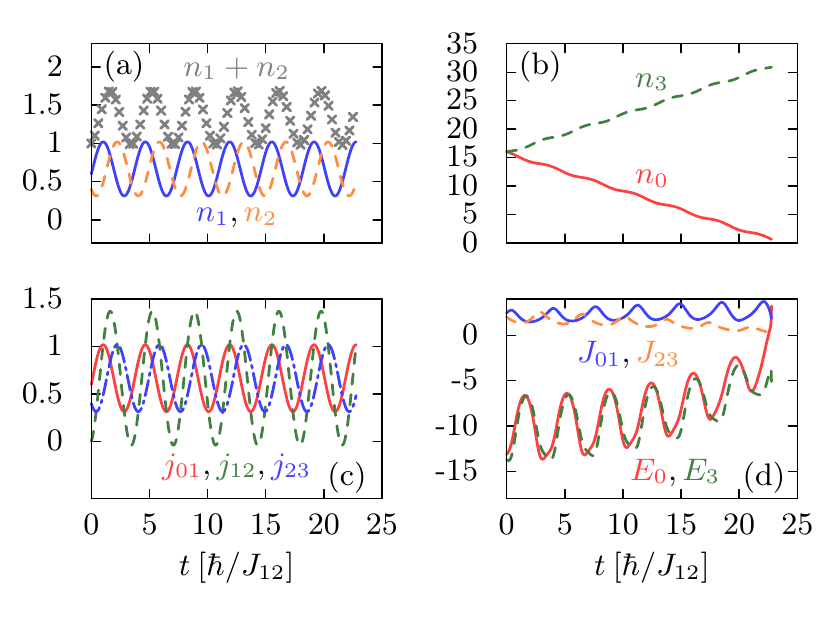}
  \caption{Same as Fig.~\ref{fig:stat_sol}, but for oscillatory
    dynamics. Additionally, the added number of particles $n_1+n_2$ is
    plotted as crosses in (a). As the particle number in the reservoir
    well~0 decreases and gets close to zero, the conditions cannot be
    fulfilled anymore and the simulation breaks down.}
  \label{fig:dynamics}
\end{figure}

As a second example we prepare a non-stationary solution at $t=0$,
with $\psi_1(0) = \sqrt{0.6}$ and $\psi_2(0) = \sqrt{0.4}$ (see
Fig.~\ref{fig:dynamics}). In this case the particle numbers in the two
middle wells oscillate in time with a phase difference $\Delta \phi <
\pi$ leading to a non-constant added number of particles, which is a
typical feature of $\pt$-symmetric systems. Since well~0 acts as a
particle reservoir the number of particles decreases and gets close to
zero up to the point where the linear system of
equations~\eqref{eq:lgs} cannot be fulfilled anymore (determinant of
coefficient matrix equals zero). This shows that the time available
for a simulation of $\pt$ symmetry with a reservoir is limited. The
matrix elements in this case show a quasi-oscillatory behavior.

\begin{figure}[tb]
  \centering
  \includegraphics[width=\columnwidth]{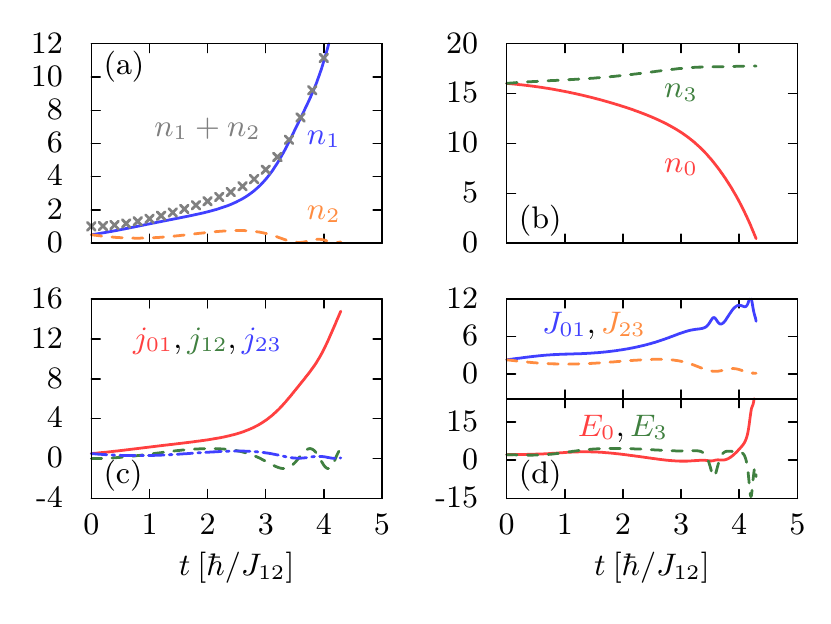}
  \caption{Same as Fig.~\ref{fig:stat_sol}, but for the interacting
    case ($c=-1$), which leads to a collapsing dynamics, indicating a
    change of stability. The reservoir is emptying in a short time
    domain such that the time available for $\pt$ symmetry is quite
    short.}
  \label{fig:collapse}
\end{figure}

Next, we consider a system with attractive interaction $c=-1$. In such
a system it is known that the ground state changes its stability in an
additional bifurcation (for repulsive interaction, the excited state
changes its stability) \cite{Haag14}. This phenomenon is highly
correlated to the occurrence of self-trapping states. In our example
$2 \times 2$ system [Eq.~\eqref{eq:ham2mode}] the ground state is
already unstable for the given interaction strength, so that the
ground state with a small perturbation is expected to collapse. The
calculation is shown in Fig.~\ref{fig:collapse}. The particle number
$n_1$ increases exponentially, which marks the beginning of
collapse. To support the exponential increase, the particle number in
the reservoir $n_0$ is decreasing exponentially, so that the
simulation breaks down after a quite short time interval. The matrix
elements now show a rather complex time-dependency.

\begin{figure}[tb]
  \centering
  \includegraphics[width=\columnwidth]{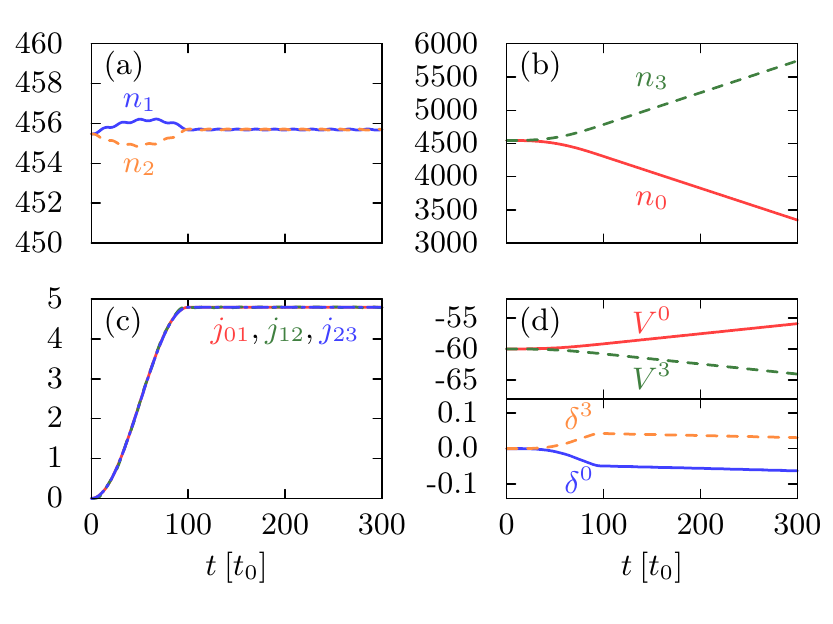}
  \caption{Same as Fig.~\ref{fig:stat_sol}, but for an adiabatic
    current ramp ($\Gamma_\text{f}/J_{12} = 0.5$, $t_\text{f}/t_0 =
    60$) for realistic parameters of the external potential. Instead
    of the matrix elements, in (d) we show the parameters of the
    potential, $\delta^k$ marks the displacement from equidistant
    potential wells. There is a total number of particles of
    $N=10^4$. See text for units.}
  \label{fig:adiab}
\end{figure}

As the last example we calculate an experimentally more realistic
scenario. Up to now -- as mentioned above -- we had to start with
appropriate wave functions in order to obtain $\pt$ symmetry. This is
nearly impossible in an experiment. To overcome this, we use an
\emph{adiabatic} change of the $\pt$ parameter from zero to its target
value. We start at the ground state of the Hermitian four-mode model
(with $\Gamma=0$) and increase $\Gamma$ to its target value
$\Gamma_\text{f}$. If this change is slow enough, we arrive
approximately at the $\pt$-symmetric ground state. The specific time
dependence we use is $\Gamma(t) = \Gamma_\text{f} [1 - \cos (\pi t /
t_\text{f})]/2$ for $t \in [0,t_\text{f}]$.

In the following we simulate a condensate of $N=10^5$ atoms of
\ce{^{87}Rb}. We use units based on the potential width in
$z$-direction, $w_z$, which we set $w_z = \unit{1}{\micro
  \meter}$. The basic unity of energy is $E_0 = \hbar^2 / m w_z^2$,
which yields $E_0/h = \unit{116}{\hertz}$. The basic unit of time is
$t_0 = m w_z^2 / \hbar = \unit{1.37}{\milli\second}$. The wells have
widths of $w_x = w_y = 4 w_z = \unit{4}{\micro\meter}$, and are
initially positioned equidistantly with a distance of $1.8 w_z =
\unit{1.8}{\micro\meter}$. The potential depths are given by $V^0=V^3
= -60 E_0$ and $V^1=V^2 = -45 E_0$. We use a scattering length of $a =
2.83 a_\text{B}$ with $a_\text{B}$ being the Bohr radius.

We calculate the adiabatic ramp for $\Gamma_\text{f}/J_{12} = 0.5$ and
$t_\text{f}/t_0 = 60$. The results are shown in
Fig.~\ref{fig:adiab}. The total particle number and currents are
scaled to a particle number of $N=10^5$, the currents are given in
units of $1/t_0$, the potential depths in units of $E_0$. Instead of
the \emph{positions} of the potential wells we plot the
\emph{difference} of the position from its initial value, $\delta^k =
s_z^k - s_z^k(t=0)$. By means of the relations derived in
Sec.~\ref{sec:from-gross-pita} we can transform from the calculated
matrix elements to parameters of the external potential.

After time $t_\text{f}$ an approximate $\pt$-symmetric state is
obtained. This can be seen by the constant number of particles in the
middle wells, despite an increasing and decreasing number in wells 3
and 0, respectively, which is a clear sign of $\pt$ symmetry, also in
a possible experiment. The potential depths of the outer wells have to
be adjusted in a simple way, the positions vary within a few percent
of the distance of the wells. This shows that the simple four-mode
model can be used to describe a realistic scenario and to obtain the
necessary parameters of the external potential.

We have applied our results from Sec.~\ref{sec:equivalence-two-four}
to different initial conditions and system parameters and have shown
that it is indeed possible to realize a $\pt$-symmetric system with a
Hermitian four-mode model. In the following Section we investigate
this scenario in the framework of the \ac{GPE} and a variational
ansatz with Gaussian functions. In particular we verify that the
results from this section agree.

\section{Variational ansatz with Gaussian functions}
\label{sec:vari-ansatz-with}

\subsection{Variational ansatz and equations of motion}
\label{sec:vari-ansatz-equat}

After the simple approach in the previous section we now investigate
the possible realization in the framework of the \ac{GPE} with an
extended variational ansatz. We use a superposition of Gaussian
functions, where -- in contrast to the simpler approach of
Sec.~\ref{sec:simpl-vari-ansatz} -- all variational parameters are
considered as time-dependent,
\begin{align}
  \label{eq:varansatz}
  \psi = \sum\limits_k \eto{- (\vec{r} - \vec{q}^k)\trans \mat{A}^k
    (\vec{r} - \vec{q}^k) + \ii \vec{p}^k (\vec{r} - \vec{q}^k) -
    \gamma^k}.
\end{align}
The matrices $\mat{A}^k = \diag( A_x^k, A_y^k, A_z^k) \in \Complex^{3
  \times 3}$ describe the width of each Gaussian function in each
direction, the vectors $\vec{q}^k = (0,0,q_z^k)\trans \in \Reals^3$
and $\vec{p}^k = (0,0,p_z^k)\trans \in \Reals^3$ the position and
momentum, respectively, and the scalar quantities $\gamma^k \in
\Complex$ the amplitude and phase. This variational ansatz can
describe a much richer dynamics than ansatz~\eqref{eq:simpleansatz}
and thus we expect more accurate results.

The equations of motion follow from the time-dependent variational
principle in the formulation of McLachlan \cite{Lachlan64}, which
states that the quantity
\begin{align}
  \label{eq:30}
  I = || \ii \hbar \phi - \hat{H} \psi ||^2
\end{align}
shall be minimized with respect to $\phi$ and $\phi = \dot{\psi}$ is
set afterwards. Details of the necessary calculation can be found in
Ref.~\cite{Eichler12}, where the ansatz~\eqref{eq:varansatz} has been
used to study the collision of anisotropic solitons in a \ac{BEC}. The
\ac{EOM} are given by
\begin{align}
  \label{eq:eom}
  \ii \hbar \sum\limits_k
  \qmprod{\firstpderiv{\psi}{z_l}}{\firstpderiv{\psi}{z_k}} \dot{z}_k
  = \matrixel{\firstpderiv{\psi}{z_l}}{\hat{H}}{\psi},
\end{align}
where $\vec{z} = (A_x^1, \dots, \gamma^\numg)\trans$ is the vector of
all variational parameters. Stationary solutions are given by fixed
points of~\eqref{eq:eom}, whereas the dynamics can be calculated by
integrating Eq.~\eqref{eq:eom} with a numerical integrator.

In Sec.~\ref{sec:few-mode-models} we started our considerations by
calculating the time-derivatives of the observables. In the case of
the continuous \ac{GPE} with a non-Hermitian potential this leads to
the modified equation of continuity
\begin{align}
  \label{eq:eoc}
  \dot{\rho} + \div \vec{j} = 2 \rho \Imag V,
\end{align}
with $\rho=\abs{\psi}^2$ and $\vec{j} = \hbar ( \psi^* \nabla \psi -
\psi \nabla \psi^*) / 2 m \ii$. The imaginary part of the potential
acts as an additional source or sink to the probability flow. Instead
of adjusting the matrix elements, we need to adjust the parameters of
the external potential. Since Eq.~\eqref{eq:eoc} is given on the whole
space $\Reals^3$, one would need to change the external potential at
every point in space, which is not manageable, neither in theory nor
in an experiment.

\begin{figure}[tb]
  \centering
  \includegraphics[width=\columnwidth]{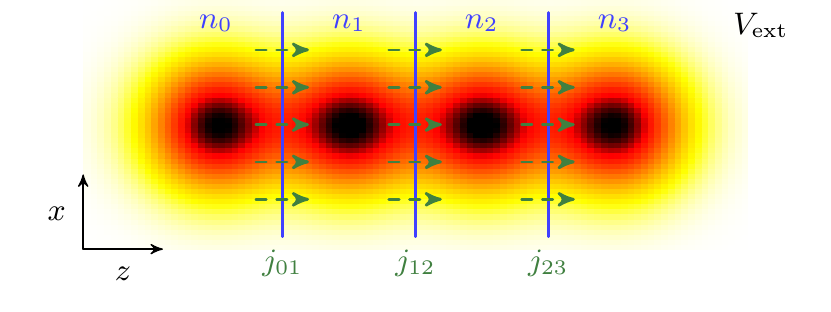}
  \caption{External potential $V_\text{ext}$; the darker, the deeper
    the potential. The wells are separated into regions by walls at
    positions right in the middle between two minima. The integral
    over $\abs{\psi}^2$ in a region $k$ is the particle number $n_k$;
    the current through a wall is denoted by $j_{k,k+1}$. See text for
    exact definitions. With this separation we obtain a formulation
    analogous to the four-mode model
    (Sec.~\ref{sec:few-mode-models}).}
  \label{fig:wells3d}
\end{figure}

To overcome this problem we use the results of
Sec.~\ref{sec:few-mode-models} as a roadmap for this continuous
system. For that, we separate the individual wells by walls between
two minima of the potential at positions $z_k=(s_z^k+s_z^{k+1})/2$ for
$k = 1,2,3$, $z_0 \to -\infty$ and $z_4 \to \infty$ (see
Fig.~\ref{fig:wells3d}). To obtain time derivatives of discrete
observables as in the few-mode model, we integrate the continuity
equation for the Hermitian four-well potential over a volume
\begin{align}
  \label{eq:31}
  V^k = \{(x,y,z)\trans | -\infty < x,y < \infty, z_{k-1} < z < z_k
  \}.
\end{align}
This yields
\begin{align}
  \label{eq:27}
  \partial_t \iiint\limits_{V^k} \dd^3 r \, \rho &= -
  \iiint\limits_{V^k} \dd^3 r \, \div \vec{j} \nn\\
  &= - \iint\limits_{\partial V^k} \dd \vec{A} \cdot \vec{j}.
\end{align}
The integral on the left-hand side can be interpreted as the number of
particles in well $k$. On the right-hand side we used Gauss's
theorem. The surface integrals goes over a box, where the areas in the
$x$-$z$- and $y$-$z$-plane go to infinity. We assume the current
$\vec{j}$ to also vanish at infinity, so only the integral over the
two surfaces in the $x$-$y$-plane contribute. Thus, we can write
\begin{align}
  \label{eq:32}
  \partial_t n_k = - \int\limits_{-\infty}^\infty \dd x
  \int\limits_{-\infty}^\infty \dd y \, \left[ j_z(x,y,z_k) -
    j_z(x,y,z_{k-1}) \right].
\end{align}
The integral over the $z$-component of $\vec{j}$ evaluated at $z_k$
represents the current from well $k$ to $k+1$, which we write as
$j_{k,k+1}$, for $z_{k-1}$ this represents the current from well $k-1$
to $k$, $j_{k-1,k}$. We finally obtain from the continuity equation
\begin{align}
  \label{eq:varobs}
  \partial_t n_k = j_{k-1,k} - j_{k,k+1},
\end{align}
where $j_{-1,0} = j_{3,4} = 0$.

Therefore, with Eq.~\eqref{eq:varobs} we have an analogous equation as
for the few-mode-model \eqref{eq:4modedt}. We can use the same method
to realize a $\pt$-symmetric system. At every time step of the
numerical integration of the equations of motion \eqref{eq:eom} we
vary the parameters of the external potential such that the
observables show the desired behavior. This is done via a nonlinear
root search, in contrast to the analytical solutions of
Sec.~\ref{sec:equivalence-two-four}. In Sec.~\ref{sec:results} we
showed that the positions of the outer wells barely needed to be
adjusted. As a simplification we only vary the depths of the outer
wells $V^0$, $V^3$ and demand the outer currents $j_{01}$, $j_{23}$ to
reach the desired values. We are left with a two-dimensional root
search. In the following section we determine the realization of $\pt$
symmetry within the \ac{GPE} and compare the results with those of the
few-mode-model.

\subsection{Results and comparison with few-mode models}
\label{sec:results-comp-with}

\begin{figure}[tb]
  \centering
  \includegraphics[width=\columnwidth]{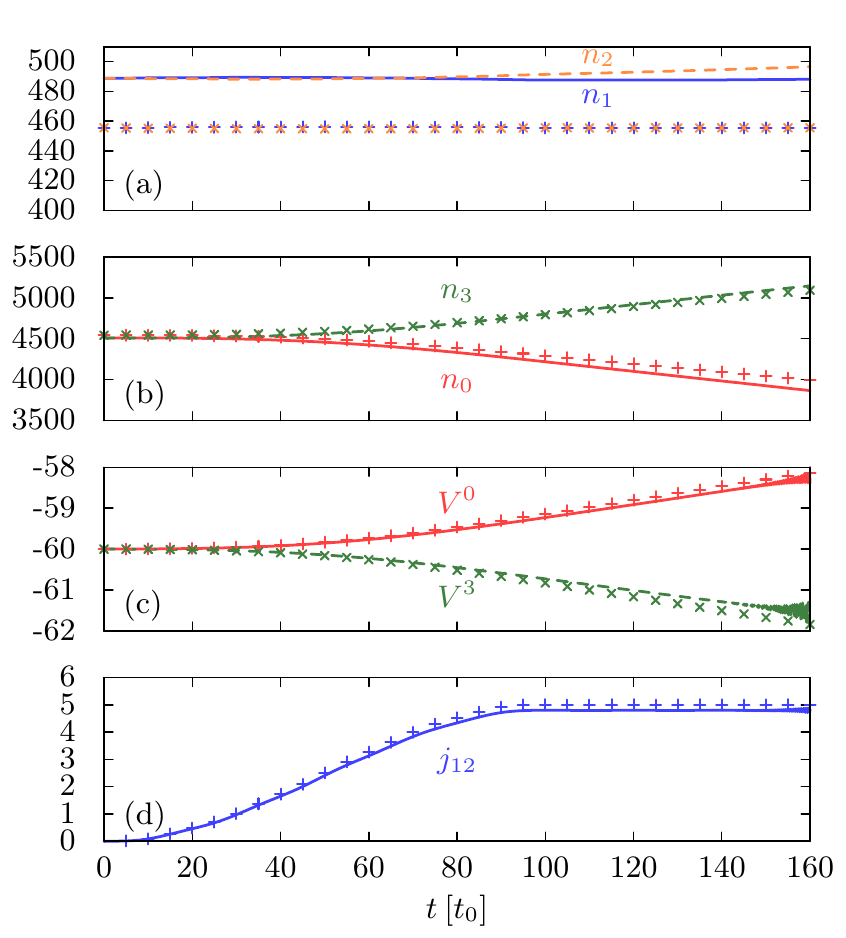}
  \caption{Adiabatic current ramp calculated with variational ansatz
    (lines) and four-mode model (crosses) for the same system. Shown
    are (a) particle numbers in the middle wells, (b) particle numbers
    in outer wells, (c) potential depths of the outer wells in units
    of $E_0$, and (d) particle current $j_{12}$. There are small
    deviations, but overall the four-mode model is a valid
    approximation compared to the variational ansatz.}
  \label{fig:compare}
\end{figure}

We use the variational ansatz to simulate the adiabatic current ramp
of Sec.~\ref{sec:results} (with same parameters) for a \ac{BEC} in a
four-well potential. For simplification, we do not give a specific
value of the $\pt$ parameter $\Gamma$ but a target particle current of
$j^\text{tar} = 5/t_0$. Fig.~\ref{fig:compare} shows the results. It
is not \emph{a priori} clear whether the quantity $n_k = (d_\eff^k)^*
d_\eff^k$ -- resulting from the \emph{transformed amplitudes} of the
simplified variational approach -- can be compared with the
\emph{integrated probability density} $n_k$ of the extended
variational approach. We show in App.~\ref{sec:prob-few-mode} that
this comparison is indeed possible and reasonable.

Overall, there is the same qualitative behavior between the two
approaches. The particle number in the middle wells is slightly under
estimated in the four-mode model due to its origin as an
approximation. Because of the fixed positions of the outer wells for
the variational approach, $n_1$ and $n_2$ are not exactly equal, but
the difference is small compared to the absolute number. We can
conclude that we can now also create an (approximate) $\pt$-symmetric
state within the variational approach and verify that the results from
the four-mode model are a good approximation.

\begin{figure}[tb]
  \centering
  \includegraphics[width=\linewidth]{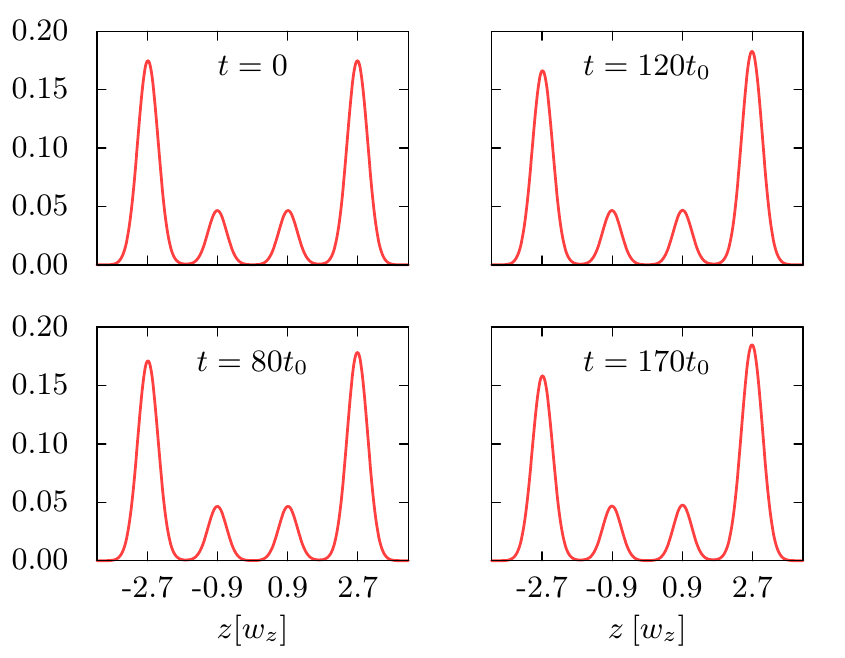}
  \caption{Modulus squared of the wave function $\abs{\psi}^2$ at
    $x,y=0$ for different times $t$. The amplitudes of the outer parts
    increase or decrease, whereas the amplitudes in the middle wells
    stay constant, a clear sign of $\pt$ symmetry, also in an
    experiment.}
  \label{fig:wavefunction}
\end{figure}

The variational approach gives us access to the wave function at each
time step (see Fig.~\ref{fig:wavefunction}). As indicated by the
particle numbers in Fig.~\ref{fig:compare}, the amplitudes in the
middle wells stay almost constant, whereas the amplitudes in the outer
wells increase or decrease in time. This could also be measured in an
experiment and could serve as a proof of realized $\pt$ symmetry.

\section{Conclusion}
\label{sec:conclusion}

By means of a simple variational ansatz -- with localized
time-independent Gaussian functions -- and the method of symmetric
orthogonalization we could transform the \ac{GPE} for a multi-well
potential to a simple few-mode model. The parameters of the
variational ansatz are obtained by an energy minimization process in a
computationally cheap way. With the results one can, on the one hand,
calculate the elements of the few-mode model for a specific system,
and on the other determine the parameters of the external potential by
knowing the time-dependent matrix elements.

These results are applied to the $\pt$-symmetric two-mode model and to
the Hermitian four-mode model. With this model we could show that the
four-mode model can be designed in such a way such that the middle
wells behave exactly as the two-wells of the $\pt$-symmetric two-mode
model. Thus, this Hermitian (closed) system can be used as a possible
realization of a quantum mechanical $\pt$-symmetric system. With an
extended Gaussian variational ansatz -- with full time-dependent
Gaussian functions, which can describe a much richer dynamics -- we
confirmed the qualitative results of the few-mode model.

In this paper we applied the idea to embed a two-well system into a
closed system for the simplest case. It is possible for future work to
use a greater embedding which could result in a simpler
time-dependence of the additional parameters. The ideas presented in
this paper could also serve as a possible road map for an
investigation of the Bose-Hubbard model and its many-particle effects.

\begin{acknowledgments}
  This work was supported by DFG\@. M.\,K.\ is grateful for support
  from the Landesgraduiertenf\"orderung of the Land
  Baden-W\"urttemberg.
\end{acknowledgments}

\appendix

\section{Analytical considerations}
\label{sec:analyt-cons}

In Sec.~\ref{sec:equivalence-two-four} we derived four conditions
[Eqs.~\eqref{eq:cond}] such that the Hermitian four-mode model can
simulate the $\pt$-symmetric two-mode model. However, only three of
them are independent, which will be shown in this appendix. For that,
we need the relations
\begin{subequations}
  \label{eq:wfrel}
  \begin{align}
    C_{jj} \tilde{j}_{ik} &= C_{ij} \tilde{j}_{jk} + \tilde{j}_{ij} C_{jk}, \\
    C_{jj} C_{ik} &= C_{ij} C_{jk} - \tilde{j}_{ij} \tilde{j}_{jk},
  \end{align}
\end{subequations}
which can be proved by simply inserting the definitions of these
quantities.

We start with Eqs.~\eqref{eq:cond1} and~\eqref{eq:cond2}, insert the
solutions~\eqref{eq:jsol}, take the square and use
Eqs.~\eqref{eq:wfrel}. We get
\begin{subequations}
  \begin{align}
    \label{eq:34}
    \tilde{j}_{01}^2 \tilde{j}_{23}^2 + \frac{2 \Gamma}{d}
    \tilde{j}_{12} \tilde{j}_{01} \tilde{j}_{23} - C_{12}^2
    \frac{n_3}{n_1} \tilde{j}_{01}^2 + \frac{4 \Gamma^2}{d^2} n_1 n_2
    &= 0, \\
    \tilde{j}_{01}^2 \tilde{j}_{23}^2 + \frac{2 \Gamma}{d}
    \tilde{j}_{12} \tilde{j}_{01} \tilde{j}_{23} - C_{12}^2
    \frac{n_0}{n_2} \tilde{j}_{23}^2 + \frac{4 \Gamma^2}{d^2} n_1 n_2
    &= 0.
  \end{align}
\end{subequations}
Taking the difference yields
\begin{align}
  \label{eq:35}
  \tilde{j}_{23}^2 &= \frac{n_2 n_3}{n_0 n_1} \tilde{j}_{01}^2 \nn\\
  \Leftrightarrow \tilde{j}_{23} &= s_3 \sqrt{\frac{n_2 n_3}{n_0 n_1}}
  \tilde{j}_{01},
\end{align}
where $s_3 = \pm 1$ gives the sign of $\tilde{j}_{23}$ compared to
$\tilde{j}_{01}$. Inserting this result into Eq.~\eqref{eq:34} gives a
fourth-order polynomial equation for $\tilde{j}_{01}$. The result is
\begin{align}
  \label{eq:36}
  \tilde{j}_{01} = s_2 \sqrt{2 n_0 n_1 \left( 1 - \alpha + s_1
      \sqrt{(1-\alpha)^2 - \beta^2} \right)},
\end{align}
with the definitions
\begin{subequations}
  \begin{align}
    \label{eq:37}
    \alpha &= \frac{\gamma}{2} \left( \beta + \frac{\gamma}{2}
    \right), \\
    \beta &= s_3 \frac{\Gamma}{d} \frac{1}{\sqrt{n_0 n_3}}, \\
    \gamma &= \frac{\tilde{j}_{12}}{\sqrt{n_1 n_2}}.
  \end{align}
\end{subequations}
Using Eqs.~\eqref{eq:wfrel} we can find all other quantities
$\tilde{j}_{kl}$ and $C_{kl}$, especially those needed for the
investigation of the fourth condition $J_{01} \tilde{j}_{02} - J_{23}
\tilde{j}_{13} = d C_{13} \tilde{j}_{02} - d C_{02} \tilde{j}_{23}$,
\begin{subequations}
  \begin{align}
    \label{eq:38}
    C_{02} &= s_2 \sign d \sqrt{2 n_0 n_2 \left( 1 - \alpha - s_1
        \sqrt{(1-\alpha)^2 - \beta^2} \right)}, \\
    C_{13} &= s_2 \sign d \sqrt{2 n_1 n_3 \left( 1 - \alpha - s_1
        \sqrt{(1-\alpha)^2 - \beta^2} \right)}, \\
    \tilde{j}_{02} &= s_6 \sqrt{2 n_0 n_2 \left( 1 + \alpha + s_1
        \sqrt{(1-\alpha)^2 - \beta^2} \right)}, \\
    \tilde{j}_{13} &= s_6 \sqrt{2 n_1 n_3 \left( 1 + \alpha + s_1
        \sqrt{(1-\alpha)^2 - \beta^2} \right)}.
  \end{align}
\end{subequations}
All signs $s_i$ are determined by the phases of the initial wave
function. We can now calculate the fourth condition and find
\begin{align}
  \label{eq:39}
  J_{01} \tilde{j}_{02} - J_{23} \tilde{j}_{13} = 0.
\end{align}
Thus, the first three conditions of Eqs.~\eqref{eq:cond} imply the
validity of the fourth one. With the results of the matrix elements of
Sec.~\ref{sec:equivalence-two-four}, namely $J_{01}$, $J_{23}$, $E_0$
and $E_3$, we find an exact equivalence of the two- and four-mode
models also for interacting atoms. Furthermore, with the results of
this appendix, we can calculate these matrix elements, once the
quantities $n_1$, $n_2$ and $\tilde{j}_{12}$ are known.

\section{Comparison of probabilities for few-mode-model and Gaussian
  functions}
\label{sec:prob-few-mode}

As discussed in Sec.~\ref{sec:results-comp-with} it is not \emph{a
  priori} clear whether we can compare the particle numbers obtained
from the four-mode model and the extended variational approach. In
this appendix we take a closer look at both quantities.

The transformation of the amplitudes from the four-mode model,
resulting from symmetric orthogonalization, is given by the inverse of
the matrix $\mat{X}$ (cf.\ Sec.~\ref{sec:symm-orth}),
\begin{align}
  \label{eq:24}
  d_\eff^l &= \sum\limits_k (\mat{X}^{-1})_{lk} d^k \nn \\
  &= \sum\limits_k \left[ (\mat{X}^{-1})^{(0)}_{lk} +
    (\mat{X}^{-1})^{(1)}_{lk} \right] d^k.
\end{align}
The individual orders of the inverse matrix can easily be calculated,
we get
\begin{subequations}
  \begin{align}
    \label{eq:33}
    (\mat{X}^{-1})_{lk}^{(0)} &= \sqrt[4]{\frac{\pi^3}{8}}
    \frac{1}{\sqrt[4]{A_{x,R}^k A_{y,R}^k A_{z,R}^k}} \delta_{lk},
    \nn\\
    (\mat{X}^{-1})_{lk}^{(1)} &= \sqrt[4]{8 \pi^3}
    \frac{\sqrt[4]{A_{x,R}^k A_{y,R}^k A_{z,R}^k A_{x,R}^l A_{y,R}^l
        A_{z,R}^l}}{\sqrt[4]{A_{x,R}^k A_{y,R}^k A_{z,R}^k} +
      \sqrt[4]{A_{x,R}^l A_{y,R}^l A_{z,R}^l}} \nn\\
    &\quad \times \frac{c^{kl}}{\sqrt{A_x^{kl}} \sqrt{A_y^{kl}}
      \sqrt{A_z^{kl}}} (\delta_{k,l-1} + \delta_{k,l+1}).
  \end{align}
\end{subequations}
The main contribution to the particle number $n_k = (d_\eff^k)^*
d_\eff^k$ is then given by
\begin{align}
  \label{eq:partnumfm}
  n_k = \sqrt{\frac{\pi^3}{8}} \frac{1}{\sqrt{A_{x,R}^k A_{y,R}^k
      A_{z,R}^k}} \abs{d^k}^2,
\end{align}
plus terms involving neighboring amplitudes.

The extended variational ansatz is given by
\begin{align}
  \label{eq:41}
  \psi = \sum\limits_k d^k \eto{-A_x^k x^2 - A_y^k y^2 - A_z^k
    (z-q_z^k)^2 + \ii (z-q_z^k) p_z^k},
\end{align}
with the amplitude $d^k = \exp(-\gamma^k)$. The probability density
can then be written as
\begin{align}
  \label{eq:42}
  \rho = \sum\limits_{k,l} d^k (d^l)^* \eto{-A_x^{kl} x^2 - A_y^{kl}
    y^2 - A_z^{kl} z^2 + \ii \tilde{p}_z^{kl} z -
    \tilde{\gamma}^{kl}},
\end{align}
with the definitions
\begin{subequations}
  \begin{align}
    \label{eq:43}
    \tilde{p}_z^{kl} &= 2 [ A_z^k q_z^k + (A_z^l)^* q_z^l] + \ii
    (p_z^k - p_z^l), \\
    \tilde{\gamma}^{kl} &= A_z^k (q_z^k)^2 + (A_z^l)^* (q_z^l)^2 + \ii
    (q_z^k p_z^k - q_z^l p_z^l).
  \end{align}
\end{subequations}
We integrate $\rho$ over the region discussed in
Sec.~\ref{sec:vari-ansatz-equat}. The integral over $x$ and $y$ yields
\begin{align}
  \label{eq:44}
  \rho_z = \sum\limits_{k,l} d^k (d^l)^* \frac{\pi}{\sqrt{A_x^{kl}}
    \sqrt{A_y^{kl}}} \eto{-A_z^{kl} z^2 + \tilde{p}_z^{kl} z -
    \tilde{\gamma}^{kl}}.
\end{align}
The integral over $z$ goes over a finite interval, say $a \leq z \leq
b$, we can express the result in terms of the error function,
\begin{align}
  \label{eq:45}
  n(a,b) &= \frac{1}{2} \sum\limits_{k,l} d^k (d^l)^*
  \frac{\pi^3}{\sqrt{A_x^{kl}} \sqrt{A_y^{kl}} \sqrt{A_z^{kl}}}
  \eto{(\tilde{p}_z^{kl})^2/4 A_z^{kl} - \tilde{\gamma}^{kl}} \nn\\
  \quad &\times \left[ \erf \left( \frac{2 A_z^{kl} b -
        \tilde{p}_z^{kl}}{4 \sqrt{A_z^{kl}}} \right) - \erf \left(
      \frac{2 A_z^{kl} a - \tilde{p}_z^{kl}}{4 \sqrt{A_z^{kl}}}
    \right) \right].
\end{align}
To obtain the number of particles in well $i$, we evaluate this
expression at points $a$ and $b$ between the wells, we can write $a =
q_z^i - \ell / 2$ and $b = q_z^i + \ell / 2$ with $\ell$ the distance
between two wells. As in the case for the four-mode model, the main
contribution of the sum results for the summand with $i=k=l$. We get
approximately
\begin{align}
  \label{eq:partnumgauss}
  n_i = \sqrt{\frac{\pi^3}{8}} \frac{\erf \left( \sqrt{A_{z,R}^i
        \ell^2 / 2} \right)}{\sqrt{A_{x,R}^i A_{y,R}^i A_{z,R}^i}}
  \abs{d^i}^2.
\end{align}
For typical values of $A_{z,R}^i \ell^2$ the error function is close
to unity. Thus, it is reasonable to compare the particle numbers
obtained from the four-mode model [Eq.~\eqref{eq:partnumfm}] with
those from the extended variational ansatz
[Eq.~\eqref{eq:partnumgauss}]. Since the particle currents are given
by the derivatives of the particle numbers, this is also the case for
the particle currents.

\end{document}